\documentclass[]{aa}  

\usepackage{graphicx}
\usepackage{txfonts}
\usepackage{xcolor}
\definecolor{seagreen}{rgb}{0.190, 0.525, 0.361}
\definecolor{darksalmon}{rgb}{0.914, 0.588, 0.478}
\definecolor{anthracite}{rgb}{0.271, 0.270, 0.318}
\definecolor{refereebrown}{rgb}{0.11, 0.11, 0.0}

\newcommand{\REFEREE}[1]{\textcolor{refereebrown!100}{\textbf{#1}}}
\begin{document} 
\title{Introducing a new multi-particle collision method for the evolution of dense stellar systems II}
   \subtitle{Core collapse}
 \titlerunning{Star cluster CC with the MPC method}
%
\author{Pierfrancesco Di Cintio
          \inst{\star 1,2,3}
          \and
          Mario Pasquato\thanks{Equal First Authors}
          \inst{4}
          \and
        Alicia Simon-Petit
          \inst{1,2}         
          \and
          Suk-Jin Yoon\inst{5}
          }

   \institute{
         Dipartimento di Fisica e Astronomia \& CSDC, Universit\`a di Firenze, via G. Sansone 1, I--50019 Sesto Fiorentino, Italy\\
             \email{pierfrancesco.dicintio@unifi.it}\\
             \email{alicia.simonpetit@unifi.it}\\
    \and      
     INFN - Sezione di Firenze, via G. Sansone 1, I--50019 Sesto Fiorentino, Italy\\
     \and
     CREF, Via Panisperna 89A, I--00184 Rome, Italy\\
     \and
     Center for Astro, Particle and Planetary Physics (CAP$^3$), New York University Abu Dhabi\\
            \email{mp5757@nyu.edu}
                 \\
 \and 
         Department of Astronomy \& Center for Galaxy Evolution Research, Yonsei University,\\ Seoul 120-749, Republic of Korea\\
             \email{sjyoon0691@yonsei.ac.kr}\\
             }

   \date{Received September 15, 1996; accepted March 16, 1997}
  \abstract
{In a previous paper we introduced a new method for simulating collisional gravitational $N$-body systems with linear time scaling on $N$, based on the Multi-Particle Collision (MPC) approach. This allows us to easily simulate globular clusters with a realistic number of stellar particles ($10^5 - 10^6$) in a matter of hours on a typical workstation.}{We evolve star clusters containing up to $10^6$ stars to core collapse and beyond. We quantify several aspects of core collapse over multiple realizations and different parameters, while always resolving the cluster core with a realistic number of particles.}{We run a large set of $N$-body simulations with our new code MPCDSS. The cluster mass function is a pure power-law with no stellar evolution, allowing us to clearly measure the effects of the mass spectrum on core collapse.}
{Leading up to core collapse, we find a power-law relation between the size of the core and the time left to core collapse. Our simulations thus confirm the theoretical self-similar contraction picture but with a dependence on the slope of the mass function. The time of core collapse has a non-monotonic dependence on the slope, which is well fit by a parabola. This holds also for the depth of core collapse and for the dynamical friction timescale of heavy particles. Cluster density profiles at core collapse show a broken power law structure, suggesting that central cusps are a genuine feature of collapsed cores. The core bounces back after collapse, with visible fluctuations, and the inner density slope evolves to an asymptotic value. The presence of an intermediate-mass black hole inhibits core collapse, making it much shallower irrespective of the mass-function slope.}{We confirm and expand on several predictions of star cluster evolution before, during, and after core collapse. Such predictions were based on theoretical calculations or small-size direct $N$-body simulations. Here we put them to the test on MPC simulations with a much larger number of particles, allowing us to resolve the collapsing core.}  

   \maketitle
%
\section{Introduction}
Following an initial formation phase from a collapsing parent cloud \citep[][]{2019ARA&A..57..227K, 2020SSRv..216...64K}, star clusters that survive early gas expulsion \citep[see e.g.][]{2020ApJ...900L...4P} undergo a secular quasi-equilibrium evolution. An initial core contraction phase is ended by a watershed moment, when core collapse halts and reverses as binary burning begins \citep[][]{1994MNRAS.268..257G, 1994MNRAS.270..298G, 2002MNRAS.336.1069B, 2010MNRAS.408L..16G, 2012MNRAS.422.3415A}. The dynamics of the following gravothermal oscillations \citep[][]{1983MNRAS.204P..19S, 1987ApJ...313..576G, 1992MNRAS.257..245A} was found to be characterized by a low-dimensionality chaotic attractor \citep[][and references therein]{1995ApJ...448..672B}.
More generally, the origin and implications of core-collapse have historically been studied both analytically \citep[][]{1938ZaTsA..22...19A, 1940MNRAS.100..396S, 1942psd..book.....C, 1961AnAp...24..369H, 1968MNRAS.138..495L,1979MNRAS.186..155H,1979MNRAS.188..525H} and through simulations \citep[][]{1970MNRAS.147..323L, 1975ApJ...201..773S, 1975IAUS...69..133H,2003MNRAS.341..247B}; see \cite{1997A&ARv...8....1M} for an early review of both. 
After the importance of core-collapse had been established, a large body of work was carried out on the subject of star-cluster evolution all the way to core-collapse and beyond, strongly relying on direct $N$-body simulations with $N$ increasing over the years as hardware capabilities improved \citep[see e.g.][]{1996MNRAS.282...19S, 1996ApJ...471..796M, 2003MNRAS.340..227B, 2010ApJ...708.1598T, 2012MNRAS.425.2872H, 2012MNRAS.427..167S, 2014MNRAS.445.3435H}. The time complexity of direct $N$-body is, however, at least quadratic \citep[e.g.][]{1999PASP..111.1333A, 2007NewA...12..357H}. Thanks to parallel codes running on GPUs \citep[][]{2015MNRAS.450.4070W}, current simulations reach $N \approx 10^6$, but this still requires several thousands of hours on a dedicated computer cluster \REFEREE{(\citealt{2016MNRAS.458.1450W}; see also e.g. \citealt{doi:10.1142/9789814374774_0009} and references therein)}, \REFEREE{with very recent N-body codes mitigating this issue \citep[][]{2020MNRAS.497..536W} while still not changing the overall scaling behaviour.} This makes replication of any given numerical experiment impractical and forces researchers to rely at best on just a few realizations of a given system.\\
\indent While open clusters can essentially be simulated with a $1$:$1$ ratio between real stars and simulation particles, the fact that large direct $N$-body simulations are impractical has detrimental implications for modeling globular clusters (except perhaps the smaller ones) and larger systems such as nuclear star clusters. It was realized very early \citep[][]{1987ApJ...313..576G} that tacitly assuming that we can scale up the results of small direct $N$-body simulations by one or more orders of magnitude is dangerous, as post-collapse dynamical behaviour can become qualitatively different with increasing numbers of particles. In addition, even if the rescaled crossing times are equal, direct simulations with substantially different number of particles have significantly different effective two-body relaxation times, that explicitly depend on the number of particles $N$ as $N/\log(N)$. As it has been already recognized in the context of Cosmological simulations (e.g. see \citealt{2002MNRAS.333..378B,2004MNRAS.348..977D,2006MNRAS.370.1247E}) this might lead to big differences in the end states of two simulations representing the ''same" system of total mass $M$ with different numbers of particles. In particular, all instability processes associated to discreteness effects will set up at earlier times (in units of a given dynamical time as function of a fixed mean mass density $\bar\rho$, $t_{\rm dyn}=1/\sqrt{G\bar\rho}$) when a smaller number of particles is used in the simulation.\\
\indent Moreover, low-$N$ simulations including an intermediate-mass black hole of mass $M_{\mathrm{IMBH}}$ in a star cluster core of mass $M_{\mathrm{core}}$ with average stellar mass $\langle m \rangle$ are bound to be unrealistic either by underestimating the $M_{\mathrm{IMBH}}/\langle m \rangle$ ratio or overestimating the $M_{\mathrm{IMBH}}/M_{\mathrm{core}}$ ratio because $M_{\mathrm{core}}/\langle m \rangle$ is the (unrealistically low) number of stars included in the simulated core. By doing so, for example the dynamical friction time scale for a displaced black hole sinking back into the star cluster core might be significantly altered, thus leading to potentially wrong conclusions on the dynamics of the cluster itself. A smaller number of simulation particles at fixed total cluster mass, also affects the formation of the loss-cone, as the latter is primarily governed by the mass ratio between the black hole and the stars.\\
\indent An alternative to direct $N$-body simulations are approximate methods, typically based on solving the Fokker-Planck equation \citep[e.g.][for a state-of-the-art Montecarlo solver]{2013MNRAS.429.1221H}, resulting in dramatically shorter run-times. In a previous paper \citep[][henceforth D2020]{2020arXiv200616018D} we introduced a code for simulating gravitational $N$-body systems which takes a new approach to approximating collisional evolution through the so-called multi-particle collision method. We refer the reader to D2020 for details on the method, its rationale and its implementation. In this work we focus on using our code to simulate star clusters containing up to $10^6$ particles through core collapse, calculating several indicators of the cluster's dynamical state and comparing with theoretical expectations. Because our typical simulation takes no more than a few hours on an ordinary workstation, we can run multiple realizations of any given system and explore the relevant parameter space at leisure, in particular exploring the effect of varying the mass-function slope. In order to achieve additional speed-up and simplify the numerical simulations, in the implementation of MPCDSS used here we neglect stellar evolution\footnote{We note that, including stellar evolution in a particle-based simulator for GCs costs only a order $N$ increase in computational time, the bottleneck of the integration being always the evaluation of the gravitational forces, scaling at best as $N\log(N)$.}, both single and binary. While for now we do not model binaries, they will be included in an upcoming version of the code (Di Cintio et al. in prep.). While these choices reduce the numerical complexity of the simulations at the expense of some loss of realism, they also allow us to disentangle the purely dynamical causes of several specific phenomena from those due to, for example, stellar evolution; this facilitates comparisons with purely theoretical works.
\section{Simulations}
\label{simulazij}
\subsection{Initial conditions}
We run a set of hybrid particle-mesh-multiparticle collision simulations using the newly introduced MPCDSS code (in D2020 we compared this method to direct $N$-body simulations, showing similar results despite dramatically shorter runtimes) on a 8 core workstation. The number of simulation particles ranges from $10^4$ to $10^6$, initially distributed following the \cite{1911MNRAS..71..460P} profile
\begin{equation}\label{plummer}
\rho(r)=\frac{3}{4\pi}\frac{Mr_s^2}{(r_s^2+r^2)^{5/2}},
\end{equation}
of total mass $M$ and scale radius $r_s$. The mass function is a pure power-law mass of the form 
\begin{equation}\label{mspectrum}
\mathcal{F}(m) = \frac{C}{m^{\alpha}};\quad m_{\rm min} \leq m\leq m_{\rm max}, 
\end{equation}
of which \cite{1955ApJ...121..161S} is a special case corresponding to $\alpha=2.3$, and where the normalization constant $C$ depends on the minimum-to-maximum-mass ratio $\mathcal{R}=m_{\rm min}/m_{\rm max}$ so that $\int_{m_{\rm min}}^{m_{\rm max}}\mathcal{F}(m){\rm d}m=M$.\\
\indent In the simulations presented in this work we concentrate on the three values of $\mathcal{R}=10^{-2}$, $10^{-3}$ and $10^{-4}$. The exponent $\alpha$ spans from $0.6$ to $3.0$ in increments of $0.1$. The systems evolve in isolation, stellar evolution is turned off, and the primordial binary fraction is set always to zero.
\begin{table}
\caption{Parameters of the initial conditions for our runs. \label{table:alpha}} 
\begin{tabular}{llll}
\hline
$N$ & $\alpha$  & $\mathcal{R}$ & $M_{\rm IMBH}$ \\
\hline \hline
$2\times10^5$ & 0.6 & $10^{-4}$, $10^{-3}$, $10^{-2}$ & $-$ \\
$2\times10^5$ & 0.7 & $10^{-4}$, $10^{-3}$, $10^{-2}$ & $-$ \\
$2\times10^5$ & 0.8 & $10^{-4}$, $10^{-3}$, $10^{-2}$ & $-$ \\
$2\times10^5$ & 0.9 & $10^{-4}$, $10^{-3}$, $10^{-2}$ & $-$ \\
$2\times10^5$ & 1.0 & $10^{-4}$, $10^{-3}$, $10^{-2}$ & $-$ \\
$2\times10^5$ & 1.1 & $10^{-4}$, $10^{-3}$, $10^{-2}$ & $-$ \\
$2\times10^5$ & 1.2 & $10^{-4}$, $10^{-3}$, $10^{-2}$ & $-$ \\
$2\times10^5$ & 1.3 & $10^{-4}$, $10^{-3}$, $10^{-2}$ & $-$ \\
$2\times10^5$ & 1.4 & $10^{-4}$, $10^{-3}$, $10^{-2}$ & $-$ \\
$2\times10^5$ & 1.5 & $10^{-4}$, $10^{-3}$, $10^{-2}$ & $-$ \\
$2\times10^5$ & 1.6 & $10^{-4}$, $10^{-3}$, $10^{-2}$ & $-$ \\
$2\times10^5$ & 1.7 & $10^{-4}$, $10^{-3}$, $10^{-2}$ & $-$ \\
$2\times10^5$ & 1.8 & $10^{-4}$, $10^{-3}$, $10^{-2}$ & $-$ \\
$2\times10^5$ & 1.9 & $10^{-4}$, $10^{-3}$, $10^{-2}$ & $-$ \\
$2\times10^5$ & 2.0 & $10^{-4}$, $10^{-3}$, $10^{-2}$ & $-$ \\
$2\times10^5$ & 2.1 & $10^{-4}$, $10^{-3}$, $10^{-2}$ & $-$ \\
$2\times10^5$ & 2.2 & $10^{-4}$, $10^{-3}$, $10^{-2}$ & $-$ \\
$2\times10^5$ & 2.3 & $10^{-4}$, $10^{-3}$, $10^{-2}$ & $-$ \\
$2\times10^5$ & 2.4 & $10^{-4}$, $10^{-3}$, $10^{-2}$ & $-$ \\
$2\times10^5$ & 2.5 & $10^{-4}$, $10^{-3}$, $10^{-2}$ & $-$ \\
$2\times10^5$ & 2.6 & $10^{-4}$, $10^{-3}$, $10^{-2}$ & $-$ \\
$2\times10^5$ & 2.7 & $10^{-4}$, $10^{-3}$, $10^{-2}$ & $-$ \\
$2\times10^5$ & 2.8 & $10^{-4}$, $10^{-3}$, $10^{-2}$ & $-$ \\
$2\times10^5$ & 2.9 & $10^{-4}$, $10^{-3}$, $10^{-2}$ & $-$ \\
$2\times10^5$ & 3.0 & $10^{-4}$, $10^{-3}$, $10^{-2}$ & $-$ \\
$10^6$ & 2.3 & $10^{-2}$ & $-$ \\
$10^6$ & 2.3 & $10^{-2}$ & $3\times 10^{-4}$ \\
$10^6$ & 2.3 & $10^{-2}$ & $10^{-3}$ \\
$10^6$ & 2.3 & $10^{-2}$ & $3\times 10^{-3}$ \\
\hline
\end{tabular}
\end{table}
\subsection{Numerical scheme}
In line with D2020 we evolved all sets of simulations for about $10^4$ dynamical times $t_{\rm dyn}\equiv\sqrt{r_s^3/GM}$, so that in all cases the systems reach core collapse and are evolved further after it for at least another $10^3t_{\rm dyn}$. We employed our recent implementation of MPCDSS where the collective gravitational potential and force are computed by standard particle-in-cell schemes on a fixed spherical grid
of $N_g=N_r\times N_\vartheta\times N_\varphi$ mesh points, while the inter-particle (dynamical) collisions are resolved using the so-called multiparticle collision scheme (hereafter MPC, see \citealt{1999JChPh.110.8605M}).\\
\indent In the simulations presented here we have solved the Poisson equation $\Delta\Phi=-4\pi G\rho$  using a fixed mesh with $N_r=1024$, $N_\vartheta=16$ and $N_\varphi=16$ with logarithmically spaced radial bins extended up to $100r_s$\footnote{Particles outside that radius are influenced by the monopole $1/r$ term of the mass distribution.} with the spherical particle mesh method by \citealt{1990MNRAS.242..595L}. Since the systems under consideration are supposed to maintain their spherical symmetry we have averaged the potential $\Phi$ along the azimuthal and polar coordinates in order to reduce the small-$N$  noise in scarcely populated cells noise and enforce the spherical symmetry throughout the simulation.\\
\indent The MPC (see \citealt{2017PhRvE..95d3203D}, D2020 for further details) essentially consist of a cell-dependant rotation of particle velocities in the cell's centre of mass frame moving at $\mathbf{u}_{\rm com}$ in the simulation's frame, that for the $j-$th particle of velocity $\mathbf{v}_j$ in cell $i$ reads 
\begin{equation}\label{rotation}
\mathbf{v}_{j}^\prime=\mathbf{u}_i+\delta\mathbf{v}_{j,\perp}{\rm cos}(\alpha_i)+(\delta\mathbf{v}_{j,\perp}\times\mathbf{R}_i){\rm sin}(\alpha_i)+\delta\mathbf{v}_{j,\parallel}.
\end{equation}  
In the equation above $\mathbf{R}_i$ is a random rotation axis, $\delta\mathbf{v}_j=\mathbf{v}_j-\mathbf{u}_i$ and $\delta\mathbf{v}_{j,\perp}$ and $\delta\mathbf{v}_{j,\parallel}$ are the relative velocity components perpendicular and parallel to $\mathbf{R}_i$, respectively. The rotation angle $\alpha_i$ when chosen randomly yields a MPC rule that only preserves Kinetic energy and momentum. If it is fixed by a particular function (again see D2020 for the details) of particles positions in the simulation frame and velocities in the centre-of-mass frame also allows one to preserve one of the components of the angular momentum vector.\\
\indent In the simulations discussed in this work, the MPC operation is performed on a different polar mesh with respect to that of the potential calculation, with $N_g=32\times 16\times 16$ points,  extended only up to $r_{\rm cut}=20r_s$. By doing so, particles at larger radii (thus in a region where the density is extremely low) are not affected by collisions.\\
\indent In all simulations presented here we use the same normalization such that $G=M=r_s=t_{\rm dyn}=1$. In these units we adopt a constant times step $\Delta t=10^{-2}$ that always assures a good balance between accuracy and computational cost.\\
\indent In the runs including the central IMBH, its interaction with the stars is evaluated directly, i.e. the IMBH does not take part in the MPC step nor in the evaluation of the mean field potential. In order to keep the same rather large $\Delta t$ of the simulation, the potential exerted by the IMBH is smoothed as $\Phi_{\rm IMBH}=-GM_{\rm IMBH}/\sqrt{r^2+\epsilon^2}$ where we take $\epsilon=10^{-4}$ in units of $r_s$ so that for the IMBH mass-to-cluster mass ratio $10^{-3}$, the softening length is always of the order of one tenth of the influence radius of the IMBH.
\section{Results}
\begin{figure}
\includegraphics[width = 0.95\columnwidth]{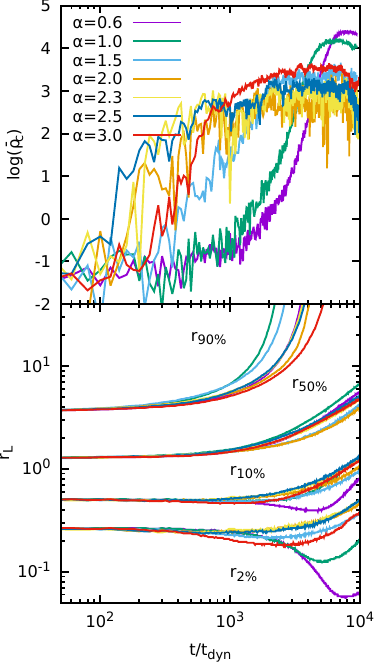}
\caption{Core density $\bar\rho_c$ (upper panel) as a function of time in units of the simulation's dynamical timescale. Evolution of 3D Lagrangian radii in units of the initial Plummer scale radius $r_s$ (from bottom to top $2\%$, $10\%$, $50\%$ and $90\%$; lower panel). Simulations with mass function slope $\alpha = 0.6$, $1.0$, $1.5$, $2.0$, $2.3$, $2.5$ and $3.0$ and $\mathcal{R}=10^{-3}$ are shown.\label{f01}}
\end{figure}
\begin{figure}
\includegraphics[width = 0.95\columnwidth]{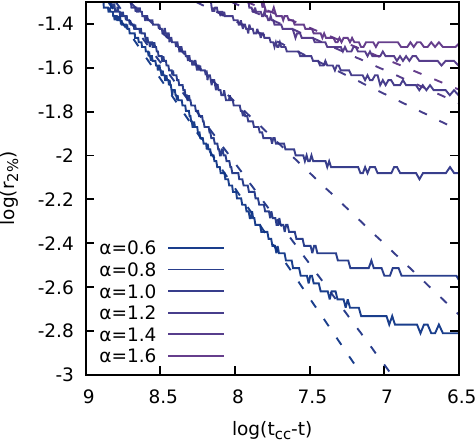}
\caption{Power-law dependence between core size ($r_{2\%}$ 3D Lagrangian radius) and time left to core collapse $t_{\mathrm{cc}} - t$. A power-law relation (appearing linear in log-log scale) holds in the initial phases of core collapse. We show it here for models with mass function slope $\alpha = 0.6$, $0.8$, $1.0$, $1.2$, $1.4$, and $1.6$, number of stars $N=2\times 10^5$ and $\mathcal{R}=10^{-3}$. Due to our definition of the x axis, time increases from the right to the left. The solid lines are data from our simulations, the superimposed dashed lines are a robust linear fit between the initial time and $t_{cc} - 100t_{\rm dyn}$. The angular coefficient of the regression lines appears to vary systematically with the mass function. As expected, the power-law relation breaks before core-collapse, when the self-similar contraction phase ends.\label{f01.5}}
\end{figure}
\begin{figure}
\includegraphics[width = 0.95\columnwidth]{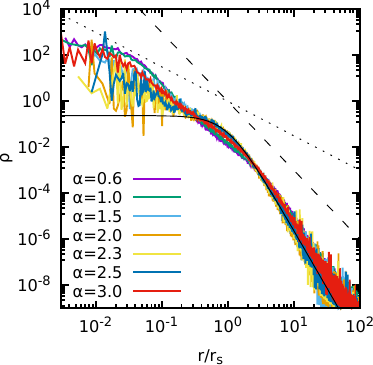}
\caption{Double power-law 3D density profile at $t_{cc}$ for models with $\alpha=0.6$, $1.0$, $1.5$, $2.0$, $2.3$, $2.5$ and $3.0$ and $N=2\times 10^5$, $\mathcal{R}=10^{-3}$. The initial isotropic Plummer profile is shown as a thin black solid line. The dashed and dotted lines mark the inner and outer limit trends ($\rho\sim r^{-1.5}$ and $\sim r^{-3}$\label{figrho} of the core density, respectively).}
\end{figure}
\begin{figure}
\includegraphics[width = 0.9\columnwidth]{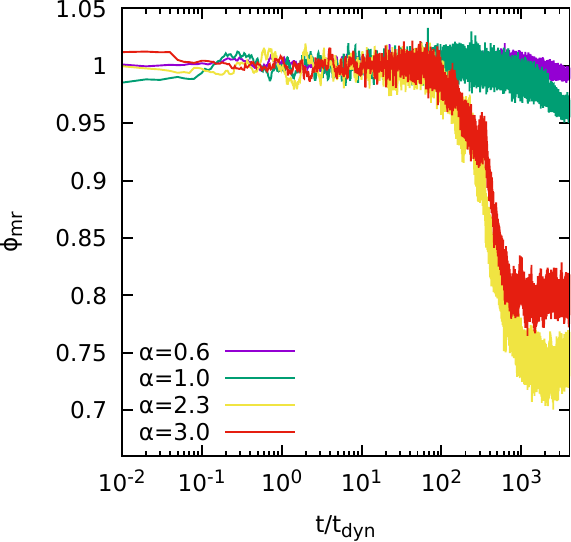}
\caption{Time evolution of the mass segregation indicator $\phi_{mr}$ computed within $r_{2\%}$ for models with $\alpha=0.6$, 1.0, 2.3 and 3.0 and $\mathcal{R}=10^{-3}$.\label{figphi}}
\end{figure}
\begin{figure}
\includegraphics[width = 0.9\columnwidth]{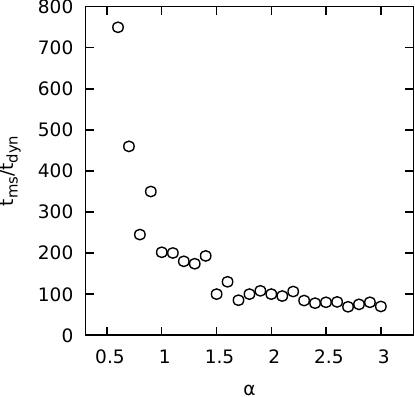}
\caption{Time of the mass segregation onset $t_{\rm ms}$ in units of $t_{\rm dyn}$, as function of the mass specrum slope $\alpha$ for the models with $\mathcal{R}=10^{-3}$.\label{figseg}}
\end{figure}
\begin{figure}
\includegraphics[width = 0.95\columnwidth]{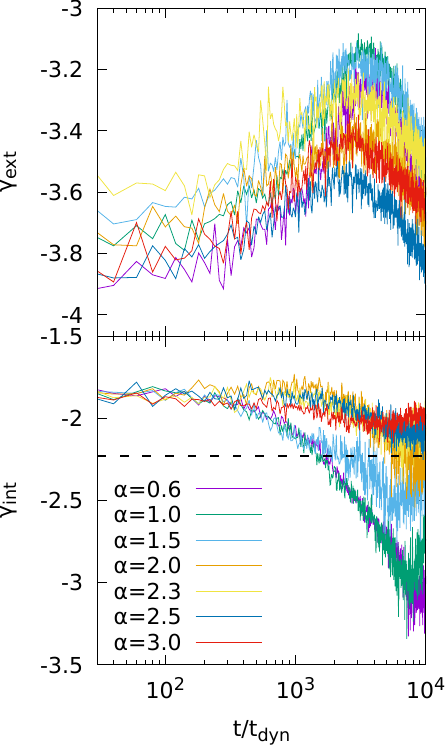}
\caption{Evolution of the 3D density slope $\gamma$ between $r_{50\%}$ and $r_{80\%}$ (top panel), and between $r_{10\%}$ and $r_{50\%}$ (bottom panel) for the models with $\alpha=0.6$, $1.0$, $1.5$, $2.0$, $2.3$, $2.5$ and $3.0$, and $N=2\times 10^5$ and $\mathcal{R}=10^{-3}$. The horizontal dashed line in the bottom panel marks the asymptotic slope $\gamma=-2.23$. \label{figgamma}}
\end{figure}
\label{resultazij}
\subsection{Evolution before core collapse}
\label{resultazij:beforecc}
We defined the time of core collapse $t_{cc}$ as the time at which the 3D radius containing the most central  $2\%$ of the simulation's particles reaches its absolute minimum. In the following we refer to this as the Lagrangian radius $r_{2\%}$. The $2\%$ radius is small enough to track the dynamics of the innermost parts of the core, while still including enough particles to be relatively unaffected by shot noise. We also calculated the 3D Lagrangian radii $r_{L}(t)$ enclosing different fractions of the total number of particles $N$ ranging from the $2\%$ to the $90\%$. We track the evolution of our simulations towards core collapse both through these radii and the central mass density $\bar\rho_c(t)$. The latter is defined as the mean 3D mass density within $5\%$ of the scale radius of our initial Plummer model, i.e. $r_m = 0.05 r_s$.\\ 
\noindent The evolution of $\bar\rho_c$ and the selected Lagrangian radii is presented in the upper and lower panels of Fig.~\ref{f01}, respectively, for the runs with $\mathcal{R}=10^{-3}$, $N=2\times10^5$ and $\alpha=0.6$, $1.0$, $1.5$, $2.0$, $2.3$, $2.5$ and $3.0$. In all simulations the central density increases monotonically (modulo fluctuations) with time until a maximum is reached, corresponding to core collapse. The slope of the mass function determines both the time at which maximum density is reached and the characteristics of the density growth before this maximum, with $\alpha=1.5$ acting as a watershed between concave and convex evolution (in log-log scale, see Fig.~\ref{f01}). In general, in models associated with larger values of $\alpha$ the central density increases more and more rapidly and settles to a somewhat constant value after core collapse, while the inner Lagrangian radii are already re-expanding.
We compare this behaviour to a simplified, equal-mass, self-similar collapse model such as the one presented in \citealt{1987degc.book.....S}, chapter 3.1 (but see also \citealt{1980MNRAS.191..483L} and the following discussion), which predicts a monotonic growth of density with time, generally with positive curvature. Clearly the presence of a mass spectrum in our simulations introduces an additional degree of freedom, complicating the behavior of the system\footnote{Actually in D2020 (Fig. 6) we have shown that the cumulative number of escapers is a largely linear function of time, irrespective of the slope of the mass function $\alpha$. Under this condition the only free parameter $\zeta$ in the model presented by chapter 3.1 of \cite{1987degc.book.....S} is fully determined, yielding a constant density as a function of time. In particular Eq. 3.6 of \cite{1987degc.book.....S} sets $\zeta = 5/3$ so the power law exponent in Eq. 3.8 becomes $0$.}. In particular, the presence of a mass spectrum implies that while the core collapse sets in, the system is also undergoing mass segregation. The latter happens at different rates depending on the structure of the mass spectrum itself, as different masses are might in principle have different dynamical friction time scales. In fact, \cite{2010AIPC.1242..117C,ciotti_2021} found that in (infinitely extended) models with exponential or power-law mass spectra the strength of dynamical friction coefficient $\nu$ is heavily affected by the  mass distribution, for test particles with masses comparable with the mean mass $\langle m\rangle$, being larger up to a factor 10 with respect to the classical case.\\
\indent All of our simulations undergo core collapse within at most four initial two-body relaxation times (defined as $t_{2b}=0.138Nt_{\rm dyn}/\log N$). From Fig.~\ref{f01} it appears that the evolutionary paths of Lagrangian radii are similar across various realizations with same $\mathcal{R}$ and $N$ but different $\alpha$. In other words, our simulated star clusters expand in average size monotonically with relatively little dependence on the mass spectrum. The latter instead has a clear influence on the evolution of the core as described by the innermost Lagrangian radii, which contract with noticeably different patterns for different $\alpha$s. \cite{1980MNRAS.191..483L} calculated analytically the time evolution of a cluster's core radius $r_c$ in the phases leading to core collapse, within the context of a self-similar collapse scenario:
\begin{equation}
\label{anali}
r_c \propto {(t_{cc} - t)}^{2/(6-\mu)}
\end{equation}
where $2 < \mu < 2.5$.
\begin{figure}
\includegraphics[width = 0.9\columnwidth]{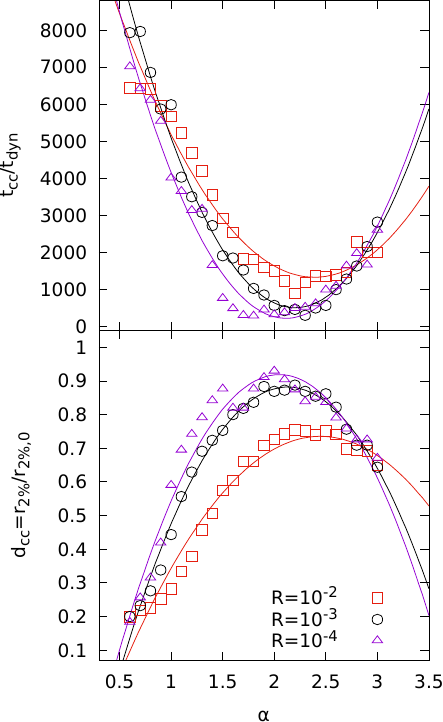}
\caption{Time of core collapse as a function of the mass function slope (top panel) and depth of core collapse (bottom panel) for isolated Plummer models with $N=2\times 10^5$, $\mathcal{R}=10^{-2}$ (red squares), $10^{-3}$ (black circles), $10^{-4}$ (purple triangles). The times are given in units of the dynamical time $t_{\rm dyn}$. The solid lines mark our second-order polynomial best fit.\label{fig3}}
\end{figure}
Our simulations are in qualitative agreement with the analytical predictions of \cite{1980MNRAS.191..483L} in the initial phase of core collapse, independently of the specific number of particles and mass ratios for $\alpha\lesssim 1.5$. In Fig.~\ref{f01.5} we show the power-law dependence of the $2\%$ Lagrangian radius on the time left to core collapse ($t_{cc} - t$) for $\alpha = 0.6, 0.8, 1.0, 1.2, 1.4, 1.6$ with $N=2\times 10^5$ and $\mathcal{R}=10^{-3}$, which appears as a linear relation in our log-log plot.\\ 
\indent For higher values of $\alpha$, for which the core collapse happens over a few hundred of $t_{\rm dyn}$, it becomes harder to fit a linear relation due to insufficient data points. Departure from \cite{1980MNRAS.191..483L} is expected to happen in the late stages of core collapse due to energy generation mechanisms in the cluster's core; this is indeed observed in Fig.~\ref{f01.5}, which shows a proportionality between $\log r_{2\%}$ and $\log {(t_{cc} - t)}$ up until $\log r_{2\%}$ saturates to a constant value as the core stops contracting.\\
\indent However, the slope of this power-law relation depends on the initial mass function of our simulation, so Eq.~\ref{anali} cannot hold with a constant $\mu$ over our whole set of simulations. This discrepancy is not surprising given that \cite{1980MNRAS.191..483L} used an equal-mass approximation. The mass spectrum appears to have a twofold effect on the dynamics of core collapse. First, for fixed total mass and density profile, it affects the rate at which the (quasi) self similar contraction happens, and second it dictates different times at which such contraction departs from self similarity (e.g. see \citealt{2007MNRAS.374..703K}). In particular, models with shallower mass spectra (i.e. smaller $\alpha$) have much longer self-similar collapse phases. We interpret this fact as an effect of the competition between mass segregation and core collapse itself.\\
\indent In order to investigate the process of mass segregation in a multi-mass system we have computed time dependently the indicator $\phi_{mr}$ defined as
\begin{equation}\label{indicator}
\phi_{mr}=\frac{\langle mr\rangle}{\langle m\rangle\langle r\rangle}.
\end{equation}
In the equation above $\langle x\rangle$ is the mean value of the quantity $x$ within a given Lagrangian radius. This indicator can be understood as the mass-weighted mean radius divided by the mean radius. Since all initial conditions used in this work are characterized by position independent mass spectra, $\phi_{mr}=1$ at all radii at $t=0$. As the systems evolve, with heavier stars sinking at smaller radii $r$, we expect $\phi_{mr}$ to decrease (at least when computed for Lagrangian radii smaller than roughly $r_{50\%}$).\\
\indent We have evaluated $\phi_{mr}$ inside different Lagrangian radii finding that the models remain essentially non-segregated up to roughly $100 t_{\rm dyn}$ while at later times, with different rates, the energy exchanges mediated by the collisions lead to a concentration of heavier stars in the inner region of the systems. As an example, in Fig. \ref{figphi} we show the time evolution of the segregation indicator for the $\alpha=0.6$, 1.0, 2.3 and 3.0 cases with $\mathcal{R}=10^{-3}$. Remarkably in all cases, $\phi_{mr}$ settles to a reasonably constant value for a few thousands dynamical times before changing appreciably, while in the same time window the corresponding Lagrangian radii are already rapidly re-expanding, compare with Fig. \ref{f01} above.\\
\indent In Fig. \ref{figseg} we show instead the time $t_{\rm ms}$ at which the mass segregation starts to take place (i.e. when $\phi_{mr}$ starts to depart sensibly from 1) as a function of the mass spectrum exponent $\alpha$. We find that such mass segregation time scale decreases monotonically for increasing $\alpha$ while the final value of $\phi_{mr}$ has a non-monotonic trend with $\alpha$.
\subsection{Density profile at core-collapse: broken power law}
\label{resultazij:atcc}
Starting from a flat-cored Plummer initial condition, the functional form of the density profile undergoes in all cases a dramatic evolution before and after the core collapse. For all explored values of $\alpha$ and $\mathcal{R}$, the density profile at the time of core collapse $t_{cc}$ presents a multiple power-law structure, as shown in Fig.~\ref{figrho} for $\alpha=0.6$, $1.0$, $1.5$, $2.0$, $2.3$, $2.5$ and $3.0$, and $\mathcal{R}=10^{-3}$. Such a multiple power-law structure is observed also in some Galactic globular clusters for which cores were resolved using the Hubble Space Telescope \citep[e.g.][]{2007AJ....134..912N}, and is often regarded as an indication of core-collapse \citep[see][for a discussion based on direct $N$-body simulations]{2010ApJ...708.1598T, 2010ApJ...720L.179V}. Typically, lower values of $\alpha$ are associated to steeper central density profiles, with both core and outer density slopes $\gamma_{\rm int}$ and $\gamma_{\rm ext}$, between $-1.5$ and $-3$.\\
\indent In Fig.~\ref{figgamma} we show that, remarkably, the evolution of the density profile for $\alpha\gtrsim 1.5$ leads at later times to a central density slope compatible with $\gamma\sim -2.23$. This holds independently of the specific value of $\alpha$, and is in agreement with what found by \cite{2012MNRAS.425.2872H, 2013MNRAS.431.2184G,2018A&A...620A..70P}, and D2020. This essentially coincides with the slope value $\gamma\sim -2.21$ found by \cite{1980MNRAS.191..483L} using a heat conduction approximation to the energy transport mediated by stellar encounters.
\subsection{Time and depth of core collapse}
\label{resultazij:cctime} 
As stated above, for each simulation we take the time of core-collapse $t_{cc}$ as the time at which the minimum value of the $r_{2\%}$ 3D Lagrangian radius is achieved. The upper panel of Fig.~\ref{fig3} shows $t_{cc}$ as a function of the initial mass-function power law $\alpha$.\\
\indent Simulations starting with a steeper mass-function (i.e. higher values of $\alpha$) are more similar to the equal-mass case, and thus reach core collapse at later times (independently of the number of particles and for all mass ratios $\mathcal{R}$, if time is measured in units of dynamical timescales of the simulation. See also Fig. 7 in D2020). This is expected by well established theory \citep[e.g.][]{1975IAUS...69....3S} showing that evolution is sped up by a mass spectrum. In addition to taking longer to reach core-collapse, low-$\alpha$ runs also reach shallower values of the core density at $t_{cc}$.\\
\indent Still, interestingly, for $\alpha>2.3$ the time of core collapse starts increasing again. The same non-monotonic trend is also observed for the depth of core collapse $d_{cc}$, defined as the ratio of $r_{2\%}$ at $t=t_{cc}$ to $r_{2\%}$ at $t=0$ (same figure, bottom panel). We fit both relations of $t_{cc}$ and $d_{cc}$ with the mass function exponent $\alpha$ using a second order polynomial (thin solid lines). As a general trend, for fixed $\alpha$ the core collapse is deeper (i.e. lower values of $d_{cc}$) and happens at later times for the models with larger minimum-to maximum mass ratio $\mathcal{R}$. This has also a similar explanation as the dependence on $\alpha$, since a larger $\mathcal{R}$ is more similar to the single-mass case.\\
\begin{figure}
\includegraphics[width = 0.9\columnwidth]{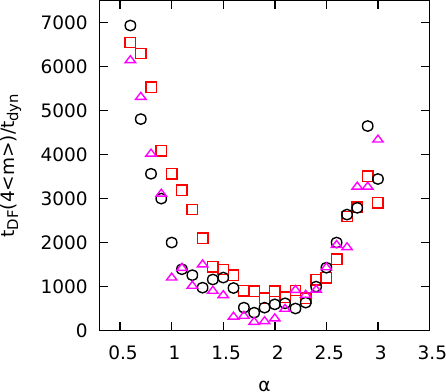}
\caption{Estimated dynamical friction time $t_{\rm DF}$ for particles with mass $4\langle m\rangle$ as function of the mass spectrum slope $\alpha$ for models with $N=2\times 10^2$ and $\mathcal{R}=10^{-2}$ (squares), $10^{-3}$ (circles) and $10^{-4}$ (triangles).\label{figdf}}
\end{figure}
\indent In Fig.~\ref{figdf} we plot the dynamical friction timescale for particles with mass $4\langle m\rangle$, showing that it is relatively unaffected by $\mathcal{R}$ while depending on $\alpha$ in a similar fashion as the timescale for core-collapse, with a minimum at intermediate $\alpha$s in the $2.0$-$2.5$ range. The lack of $\mathcal{R}$ dependence is due to the fact that changes in $\mathcal{R}$ affect only the extremes of the mass spectrum, so are not observed for particles in this relatively central mass range.
\subsection{Effects of an IMBH}
\begin{figure}
\includegraphics[width = 0.9\columnwidth]{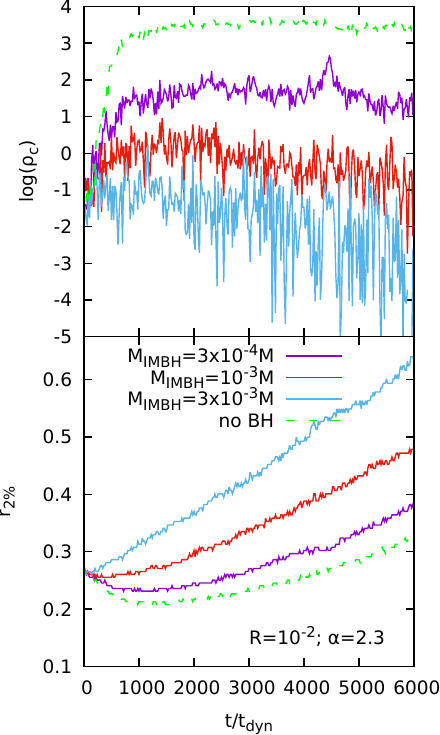}
\caption{Time evolution of the mean central density $\bar\rho_c$ (upper panel) and 3D Lagrangian radii enclosing the $2\%$ of the total number of simulation particles $N$ (lower panel) for simulations with $\alpha = 2.3$, $\mathcal{R}=10^{-3}$ with (solid lines) and without (dashed lines) IMBH. \label{figbh1}}
\end{figure}

In Fig.~\ref{figbh1} we show the evolution of central density and Lagrangian radii for three simulations containing $10^6$ particles, $\alpha = 2.3$ and an initially central IMBH of different mass, respectively $3 \times 10^{-4}$, $10^{-3}$, and $3 \times 10^{-3}$ the total simulation mass. Because $N=10^6$, the mass ratio between the IMBH and the typical star is within the correct astrophysical range for a typical globular star cluster, assuming that the mass of a IMBH ranges from $10^2$ to $10^5M_\odot$ (e.g. see \citealt{2020ARA&A..58..257G}), and the stars in globular clusters have an average mass of $\approx 0.5M_\odot$.
In all cases the core collapse stops at earlier times for the models with a central IMBH with respect to the cases without IMBH, but it is in general much shallower, involving at most a contraction of the Lagrangian radius $r_{2\%}$ of roughly $10\%$, after which the core size bounces back quickly and expands way more than in systems not hosting a central IMBH. We thus confirm the theoretical expectations that IMBHs induce swollen cores in star clusters \citep[see e.g.][and references therein]{2007MNRAS.379...93H,2008IAUS..246..351U,2012ApJ...750...31U}.
\section{Discussion and conclusions}
We used a new $N$-body simulation approach based on the multi-particle collision method to simulate star clusters with a realistic number of particles from $10^5$ up to $10^6$. We simulated systems characterized by a power-law mass function, whose exponent was varied in small increments over a wide range of values, resulting in 98 different initial conditions summarized in Table \ref{table:alpha}. This was made possible by the high performance of our code when compared to a direct $N$-body approach (with linear complexity versus quadratic in the number of particles).

We found that, for all slopes of the mass spectrum we simulated, the 3D density profile at core collapse has a broken power-law shape. This suggests that a central density cusp may be an indication of core-collapsed status in real star clusters. Thanks to the large number of particles we could simulate using our new technique, we were able to resolve this power-law cusp deep into the innermost regions of the core. Among other findings that confirm previous analytical work, we showed that the slope of this inner cusp evolves asymptotically to the value predicted by \cite{1980MNRAS.191..483L} based on an analytical model. Additionally, our simulations evolve through the initial self-similar phase of core collapse (before binary burning kicks in) following the \cite{1980MNRAS.191..483L} prediction (originally developed for a single mass system) of a power-law scaling of core radius with time to core collapse, even though the scaling law we observe has a different exponent which depends on the chosen mass spectrum. Remarkably, such a power-law behaviour is retrieved even if in the set of simulations presented here, binary formation has been neglected.\\
\indent We find a somewhat surprising parabolic dependence of the time and depth of core collapse as a function of the mass function slope. This is likely caused by the similarly non-monotonic dependence of the dynamical friction timescale on the said slope, as we plan to show analytically within the framework introduced by \cite{2010AIPC.1242..117C}.

Finally, we were able to simulate $10^6$ particle star clusters including an IMBH, thus correctly matching the $M_{\rm IMBH}/\langle m \rangle$ ratio, which is currently at the limit of direct $N$-body simulation capabilities. We confirm previous results showing that an IMBH essentially induces a faster and shallower core collapse that reverts rapidly resulting in an appreciably swollen core, with respect to models with the same initial mass distribution but without an IMBH.

\begin{acknowledgements}
This material is based upon work supported by Tamkeen under the NYU Abu Dhabi Research Institute grant CAP3. P.F.D.C. and A.S.-P. wish to thank the financing from MIUR-PRIN2017 project \textit{Coarse-grained description for non-equilibrium systems and transport phenomena
(CO-NEST)} n.201798CZL. S.-J.Y. acknowledges support by the Mid-career Researcher Program (No.2019R1A2C3006242) and the SRC Program (the Center for Galaxy Evolution
Research; No. 2017R1A5A1070354) through the National Research
Foundation of Korea. We thank the anonymous Referee for his/her comments that helped improving the presentation of our results.   
\end{acknowledgements}
   \bibliographystyle{aa} 
   \bibliography{manuscript} 

\begin{thebibliography}{61}
\expandafter\ifx\csname natexlab\endcsname\relax\def\natexlab#1{#1}\fi

\bibitem[{{Aarseth}(1999)}]{1999PASP..111.1333A}
{Aarseth}, S.~J. 1999, \pasp, 111, 1333

\bibitem[{{Alexander} \& {Gieles}(2012)}]{2012MNRAS.422.3415A}
{Alexander}, P. E.~R. \& {Gieles}, M. 2012, \mnras, 422, 3415

\bibitem[{{Allen} \& {Heggie}(1992)}]{1992MNRAS.257..245A}
{Allen}, F.~S. \& {Heggie}, D.~C. 1992, \mnras, 257, 245

\bibitem[{{Ambartsumian}(1938)}]{1938ZaTsA..22...19A}
{Ambartsumian}, V.~A. 1938, TsAGI Uchenye Zapiski, 22, 19

\bibitem[{{Baumgardt} {et~al.}(2003){Baumgardt}, {Heggie}, {Hut}, \&
  {Makino}}]{2003MNRAS.341..247B}
{Baumgardt}, H., {Heggie}, D.~C., {Hut}, P., \& {Makino}, J. 2003, \mnras, 341,
  247

\bibitem[{{Baumgardt} {et~al.}(2002){Baumgardt}, {Hut}, \&
  {Heggie}}]{2002MNRAS.336.1069B}
{Baumgardt}, H., {Hut}, P., \& {Heggie}, D.~C. 2002, \mnras, 336, 1069

\bibitem[{{Baumgardt} \& {Makino}(2003)}]{2003MNRAS.340..227B}
{Baumgardt}, H. \& {Makino}, J. 2003, \mnras, 340, 227

\bibitem[{{Binney} \& {Knebe}(2002)}]{2002MNRAS.333..378B}
{Binney}, J. \& {Knebe}, A. 2002, \mnras, 333, 378

\bibitem[{{Breeden} \& {Cohn}(1995)}]{1995ApJ...448..672B}
{Breeden}, J.~L. \& {Cohn}, H.~N. 1995, \apj, 448, 672

\bibitem[{{Chandrasekhar}(1942)}]{1942psd..book.....C}
{Chandrasekhar}, S. 1942, {Principles of stellar dynamics} (University of
  Chicago Press)

\bibitem[{{Ciotti}(2010)}]{2010AIPC.1242..117C}
{Ciotti}, L. 2010, in American Institute of Physics Conference Series, Vol.
  1242, American Institute of Physics Conference Series, ed. G.~{Bertin},
  F.~{de Luca}, G.~{Lodato}, R.~{Pozzoli}, \& M.~{Rom{\'e}}, 117--128

\bibitem[{Ciotti(2021)}]{ciotti_2021}
Ciotti, L. 2021, Introduction to Stellar Dynamics (Cambridge University Press)

\bibitem[{{Di Cintio} {et~al.}(2017){Di Cintio}, {Livi}, {Lepri}, \&
  {Ciraolo}}]{2017PhRvE..95d3203D}
{Di Cintio}, P., {Livi}, R., {Lepri}, S., \& {Ciraolo}, G. 2017, \pre, 95,
  043203

\bibitem[{{Di Cintio} {et~al.}(2021){Di Cintio}, {Pasquato}, {Kim}, \&
  {Yoon}}]{2020arXiv200616018D}
{Di Cintio}, P., {Pasquato}, M., {Kim}, H., \& {Yoon}, S.-J. 2021, \aap, 649,
  A24

\bibitem[{{Diemand} {et~al.}(2004){Diemand}, {Moore}, {Stadel}, \&
  {Kazantzidis}}]{2004MNRAS.348..977D}
{Diemand}, J., {Moore}, B., {Stadel}, J., \& {Kazantzidis}, S. 2004, \mnras,
  348, 977

\bibitem[{{El-Zant}(2006)}]{2006MNRAS.370.1247E}
{El-Zant}, A.~A. 2006, \mnras, 370, 1247

\bibitem[{{Gieles} {et~al.}(2010){Gieles}, {Baumgardt}, {Heggie}, \&
  {Lamers}}]{2010MNRAS.408L..16G}
{Gieles}, M., {Baumgardt}, H., {Heggie}, D.~C., \& {Lamers}, H. J.~G.~L.~M.
  2010, \mnras, 408, L16

\bibitem[{{Giersz} \& {Heggie}(1994{\natexlab{a}})}]{1994MNRAS.268..257G}
{Giersz}, M. \& {Heggie}, D.~C. 1994{\natexlab{a}}, \mnras, 268, 257

\bibitem[{{Giersz} \& {Heggie}(1994{\natexlab{b}})}]{1994MNRAS.270..298G}
{Giersz}, M. \& {Heggie}, D.~C. 1994{\natexlab{b}}, \mnras, 270, 298

\bibitem[{{Giersz} {et~al.}(2013){Giersz}, {Heggie}, {Hurley}, \&
  {Hypki}}]{2013MNRAS.431.2184G}
{Giersz}, M., {Heggie}, D.~C., {Hurley}, J.~R., \& {Hypki}, A. 2013, \mnras,
  431, 2184

\bibitem[{{Goodman}(1987)}]{1987ApJ...313..576G}
{Goodman}, J. 1987, \apj, 313, 576

\bibitem[{{Greene} {et~al.}(2020){Greene}, {Strader}, \&
  {Ho}}]{2020ARA&A..58..257G}
{Greene}, J.~E., {Strader}, J., \& {Ho}, L.~C. 2020, \araa, 58, 257

\bibitem[{{Harfst} {et~al.}(2007){Harfst}, {Gualandris}, {Merritt}, {Spurzem},
  {Portegies Zwart}, \& {Berczik}}]{2007NewA...12..357H}
{Harfst}, S., {Gualandris}, A., {Merritt}, D., {et~al.} 2007, \na, 12, 357

\bibitem[{{Heggie}(1979{\natexlab{a}})}]{1979MNRAS.188..525H}
{Heggie}, D.~C. 1979{\natexlab{a}}, \mnras, 76, 525

\bibitem[{{Heggie}(1979{\natexlab{b}})}]{1979MNRAS.186..155H}
{Heggie}, D.~C. 1979{\natexlab{b}}, \mnras, 186, 155

\bibitem[{Heggie(2011)}]{doi:10.1142/9789814374774_0009}
Heggie, D.~C. 2011, Problems of collisional stellar dynamics (World Scientific
  Publishing), 121--136

\bibitem[{{Heggie}(2014)}]{2014MNRAS.445.3435H}
{Heggie}, D.~C. 2014, \mnras, 445, 3435

\bibitem[{{H{\'e}non}(1961)}]{1961AnAp...24..369H}
{H{\'e}non}, M. 1961, Annales d'Astrophysique, 24, 369

\bibitem[{{H{\'e}non}(1975)}]{1975IAUS...69..133H}
{H{\'e}non}, M. 1975, in Dynamics of the Solar Systems, ed. A.~{Hayli},
  Vol.~69, 133

\bibitem[{{Hurley}(2007)}]{2007MNRAS.379...93H}
{Hurley}, J.~R. 2007, \mnras, 379, 93

\bibitem[{{Hurley} \& {Shara}(2012)}]{2012MNRAS.425.2872H}
{Hurley}, J.~R. \& {Shara}, M.~M. 2012, \mnras, 425, 2872

\bibitem[{{Hypki} \& {Giersz}(2013)}]{2013MNRAS.429.1221H}
{Hypki}, A. \& {Giersz}, M. 2013, \mnras, 429, 1221

\bibitem[{{Khalisi} {et~al.}(2007){Khalisi}, {Amaro-Seoane}, \&
  {Spurzem}}]{2007MNRAS.374..703K}
{Khalisi}, E., {Amaro-Seoane}, P., \& {Spurzem}, R. 2007, \mnras, 374, 703

\bibitem[{{Krause} {et~al.}(2020){Krause}, {Offner}, {Charbonnel}, {Gieles},
  {Klessen}, {V{\'a}zquez-Semadeni}, {Ballesteros-Paredes}, {Girichidis},
  {Kruijssen}, {Ward}, \& {Zinnecker}}]{2020SSRv..216...64K}
{Krause}, M. G.~H., {Offner}, S. S.~R., {Charbonnel}, C., {et~al.} 2020, \ssr,
  216, 64

\bibitem[{{Krumholz} {et~al.}(2019){Krumholz}, {McKee}, \&
  {Bland-Hawthorn}}]{2019ARA&A..57..227K}
{Krumholz}, M.~R., {McKee}, C.~F., \& {Bland-Hawthorn}, J. 2019, \araa, 57, 227

\bibitem[{{Larson}(1970)}]{1970MNRAS.147..323L}
{Larson}, R.~B. 1970, \mnras, 147, 323

\bibitem[{{Londrillo} \& {Messina}(1990)}]{1990MNRAS.242..595L}
{Londrillo}, P. \& {Messina}, A. 1990, \mnras, 242, 595

\bibitem[{{Lynden-Bell} \& {Eggleton}(1980)}]{1980MNRAS.191..483L}
{Lynden-Bell}, D. \& {Eggleton}, P.~P. 1980, \mnras, 191, 483

\bibitem[{{Lynden-Bell} \& {Wood}(1968)}]{1968MNRAS.138..495L}
{Lynden-Bell}, D. \& {Wood}, R. 1968, \mnras, 138, 495

\bibitem[{{Makino}(1996)}]{1996ApJ...471..796M}
{Makino}, J. 1996, \apj, 471, 796

\bibitem[{{Malevanets} \& {Kapral}(1999)}]{1999JChPh.110.8605M}
{Malevanets}, A. \& {Kapral}, R. 1999, \jcp, 110, 8605

\bibitem[{{Meylan} \& {Heggie}(1997)}]{1997A&ARv...8....1M}
{Meylan}, G. \& {Heggie}, D.~C. 1997, \aapr, 8, 1

\bibitem[{{Noyola} \& {Gebhardt}(2007)}]{2007AJ....134..912N}
{Noyola}, E. \& {Gebhardt}, K. 2007, \aj, 134, 912

\bibitem[{{Pang} {et~al.}(2020){Pang}, {Li}, {Tang}, {Pasquato}, \&
  {Kouwenhoven}}]{2020ApJ...900L...4P}
{Pang}, X., {Li}, Y., {Tang}, S.-Y., {Pasquato}, M., \& {Kouwenhoven}, M.~B.~N.
  2020, \apjl, 900, L4

\bibitem[{{Pavl{\'\i}k} \& {{\v{S}}ubr}(2018)}]{2018A&A...620A..70P}
{Pavl{\'\i}k}, V. \& {{\v{S}}ubr}, L. 2018, \aap, 620, A70

\bibitem[{{Plummer}(1911)}]{1911MNRAS..71..460P}
{Plummer}, H.~C. 1911, \mnras, 71, 460

\bibitem[{{Salpeter}(1955)}]{1955ApJ...121..161S}
{Salpeter}, E.~E. 1955, \apj, 121, 161

\bibitem[{{Sippel} {et~al.}(2012){Sippel}, {Hurley}, {Madrid}, \&
  {Harris}}]{2012MNRAS.427..167S}
{Sippel}, A.~C., {Hurley}, J.~R., {Madrid}, J.~P., \& {Harris}, W.~E. 2012,
  \mnras, 427, 167

\bibitem[{{Spitzer}(1975)}]{1975IAUS...69....3S}
{Spitzer}, L., J. 1975, in Dynamics of the Solar Systems, ed. A.~{Hayli},
  Vol.~69, 3

\bibitem[{{Spitzer} \& {Shull}(1975)}]{1975ApJ...201..773S}
{Spitzer}, L., J. \& {Shull}, J.~M. 1975, \apj, 201, 773

\bibitem[{{Spitzer}(1940)}]{1940MNRAS.100..396S}
{Spitzer}, Lyman, J. 1940, \mnras, 100, 396

\bibitem[{{Spitzer}(1987)}]{1987degc.book.....S}
{Spitzer}, L. 1987, {Dynamical evolution of globular clusters} (Princeton, NJ,
  Princeton University Press, 191 p.)

\bibitem[{{Spurzem} \& {Aarseth}(1996)}]{1996MNRAS.282...19S}
{Spurzem}, R. \& {Aarseth}, S.~J. 1996, \mnras, 282, 19

\bibitem[{{Sugimoto} \& {Bettwieser}(1983)}]{1983MNRAS.204P..19S}
{Sugimoto}, D. \& {Bettwieser}, E. 1983, \mnras, 204, 19P

\bibitem[{{Trenti} {et~al.}(2010){Trenti}, {Vesperini}, \&
  {Pasquato}}]{2010ApJ...708.1598T}
{Trenti}, M., {Vesperini}, E., \& {Pasquato}, M. 2010, \apj, 708, 1598

\bibitem[{{Umbreit} {et~al.}(2012){Umbreit}, {Fregeau}, {Chatterjee}, \&
  {Rasio}}]{2012ApJ...750...31U}
{Umbreit}, S., {Fregeau}, J.~M., {Chatterjee}, S., \& {Rasio}, F.~A. 2012,
  \apj, 750, 31

\bibitem[{{Umbreit} {et~al.}(2008){Umbreit}, {Fregeau}, \&
  {Rasio}}]{2008IAUS..246..351U}
{Umbreit}, S., {Fregeau}, J.~M., \& {Rasio}, F.~A. 2008, in Dynamical Evolution
  of Dense Stellar Systems, ed. E.~{Vesperini}, M.~{Giersz}, \& A.~{Sills},
  Vol. 246, 351--355

\bibitem[{{Vesperini} \& {Trenti}(2010)}]{2010ApJ...720L.179V}
{Vesperini}, E. \& {Trenti}, M. 2010, \apjl, 720, L179

\bibitem[{{Wang} {et~al.}(2020){Wang}, {Iwasawa}, {Nitadori}, \&
  {Makino}}]{2020MNRAS.497..536W}
{Wang}, L., {Iwasawa}, M., {Nitadori}, K., \& {Makino}, J. 2020, \mnras, 497,
  536

\bibitem[{{Wang} {et~al.}(2016){Wang}, {Spurzem}, {Aarseth}, {Giersz}, {Askar},
  {Berczik}, {Naab}, {Schadow}, \& {Kouwenhoven}}]{2016MNRAS.458.1450W}
{Wang}, L., {Spurzem}, R., {Aarseth}, S., {et~al.} 2016, \mnras, 458, 1450

\bibitem[{{Wang} {et~al.}(2015){Wang}, {Spurzem}, {Aarseth}, {Nitadori},
  {Berczik}, {Kouwenhoven}, \& {Naab}}]{2015MNRAS.450.4070W}
{Wang}, L., {Spurzem}, R., {Aarseth}, S., {et~al.} 2015, \mnras, 450, 4070

\end{thebibliography}
\end{document}